\documentclass{llncs}

\usepackage{makeidx}  
\usepackage{xcolor}
\usepackage{graphicx} 
\usepackage{caption}
\usepackage{subcaption}
\usepackage{float}
\usepackage{algpseudocode}
\usepackage{algorithm}
\usepackage{algorithmicx}
\usepackage{listings}
\usepackage{tikz} 
\usepackage{amssymb}
\usepackage{amsmath}
\usepackage{tabu} 
\usepackage{bm} 
\usepackage{booktabs} 
\usepackage{array} 
\usepackage{paralist} 
\usepackage{verbatim} 
\usepackage{subfig} 
\usepackage{bbm}

\begin{document}

\title{$LCSk$++: Practical similarity metric for long strings}
\author{Filip Paveti\'{c}\inst{1} \and Goran \v{Z}u\v{z}i\'{c}\inst{1} \and Mile \v{S}iki\'{c}\inst{1,2}}
\authorrunning{Filip Paveti\'{c} et al.}
\tocauthor{Filip Paveti\'{c}, Goran \v{Z}u\v{z}i\'{c}, Mile \v{S}iki\'{c}}

\institute{
Faculty of Electrical Engineering and Computing, University of Zagreb, Croatia\\
\and Bioinformatics Institute 30 Biopolis Street, \#07-01 Matrix, 138671 Singapore
  \email{fpavetic@gmail.com}\\
  \email{\{goran.zuzic, mile.sikic\}@fer.hr}
}

\maketitle

\begin{abstract}
In this paper we present $LCSk$++: a new metric for measuring the similarity of long strings, and provide an algorithm for its efficient computation. With ever increasing size of strings occuring in practice, e.g. large genomes of plants and animals, classic algorithms such as Longest Common Subsequence (LCS) fail due to demanding computational complexity. Recently, Benson et al. defined a similarity metric named $LCSk$. By relaxing the requirement that the $k$-length substrings should not overlap, we extend their definition into a new metric. An efficient algorithm is presented which computes $LCSk$++ with complexity of $O((|X|+|Y|)\log(|X|+|Y|))$ for strings $X$ and $Y$ under a realistic random model. The algorithm has been designed with implementation simplicity in mind. Additionally, we describe how it can be adjusted to compute $LCSk$ as well, which gives an improvement of the $O(|X|\dot|Y|)$ algorithm presented in the original $LCSk$ paper.

\begin{keywords}
Efficient longest common subsequence, Similarity of long strings, Bioinformatics, Sparse dynamic programming

\end{keywords}
\end{abstract}

\section{Introduction}

Measuring the similarity of strings is the fundamental problem which arises in many applications including DNA sequence comparison \cite{lcsk}, differential file analysis and plagiarism detection \cite{chen2010plagiarism}. Metrics such as Longest Common Subsequence \cite{lcs.hunt} or Edit Distance \cite{cormen.introduction} are usually used for solving this type of problems. Still, even advanced variants of these approaches don't cope well with long input strings (e.g. the size of the human genome).

A general approach of approximating the Longest Common Subsequence is given by Baker and Giancarlo \cite{lcsfragments}. They assume a list of matching substring pairs of various lengths as input and combine them to approximate the LCS between two long strings. The two variants of their algorithm have $O(T \log T)$ and $O(T \log \log min(T,nm/T))$ time complexities, where $T$ denotes the number of matching fragments and $n,m$ denote lengths of the two strings. We simplify some of their ideas and address the question of how to select these matching fragments.

Recently, significant effort was directed towards defining new similarity metrics. Benson et al. \cite{lcsk} defined a metric called $LCSk$, which is computed between strings $X$ of length $m$ and $Y$ of length $n$. It counts the maximal number of nonoverlapping matching $k$-length substrings\footnote{Similarly as in \cite{lcsk}, we use the term substring to denote a consecutive part of the string, while a subsequence is obtained by deleting symbols from arbitrary indices.} in the two strings (see Example \ref{example:lcsk.drawback}). An $O(mn)$ time and space algorithm is proposed for computing and reconstructing the optimal $LCSk$.  Deorowicz and Grabowski \cite{lcskfast} correctly observed that $LCSk$ can be computed more efficiently. Out of several approaches, the proposed \emph{Sparse} method allows both the computation of $LCSk$ metric and its reconstruction in $O(m + n + r \log l)$ time and $O(r)$ memory complexity, where $l$ is the length of the optimal solution and $r$ is the total number of matching $k$-length substring pairs between the input strings. In their approach they adapt the Hunt-Szymanski \cite{lcs.hunt} paradigm in a way that makes them rely on the usage of persistent red-black binary tree. 

A serious drawback to the $LCSk$ definition is that it considers only nonoverlapping matches with length of \textbf{exactly} $k$, thereby possibly ignoring substring matches with lengths of at \textbf{least} $k$ (see Example \ref{example:lcsk.drawback}). Therefore, we propose the $LCSk$++ measure, the longest common subsequence which removes this restriction. We still use the substrings of length $k$ for computation, but they are allowed to overlap which results that the resulting common subsequence consists of nonoverlapping matching substrings with length of at least\footnote{Hence the plusses in the $LCSk$++.} $k$. We give an efficient $O(m + n + r \log r)$ time complexity algorithm in section \ref{sec:efficient}. The algorithm utilizes only on a light-weight Fenwick tree \cite{fenwick1994new} data structure. In section 4 we demonstrate the ability of $LCSk$++ to separate pairs of strings which are similar from unrelated ones under a realistic random model\footnote{We call this property separability.}. We discuss the influence of the parameter $k$ on the performance and the separability. To the best of our knowledge, such discussion didn't appear in any previous related work. We conclude the paper with an $O((m+n) \log (m+n))$ time and $O(m + n)$ memory complexity algorithm with good separability under the presented random model.

\begin{example}  
Consider three strings: $X$=$ABCBA$, $Y$=$ABCBA$ and $Z$=$ABCDE$ and let $k=3$. $LCS3$ between every pair of these strings is equal to 1. The fact that $X$ and $Y$ are more similar than $X$ and $Z$ is not captured. That is because the $k$-length substrings are forbidden to overlap.
  \label{example:lcsk.drawback}
\end{example}

\section{Preliminaries}

In this section we formalize the concepts used in the remainder of the paper.

\begin{definition}[Common subsequence]
\label{def:cs}
Given two strings $X$ and $Y$ consider two sets of distinct indices $I=\{i_1, i_2, ... , i_n\}$ and $J=\{j_1, j_2, ... , j_n\}$ such that $i_1 < i_2 < ... < i_n$, $j_1 < j_2 < ... < j_n$ and $X_{i_x}=Y_{j_x}$ for $x=1...n$. Sets $I$ and $J$ determine a \textbf{common subsequence} of $X$ and $Y$ whose length is equal to $n$.
\end{definition}

\begin{definition}[k++ common subsequence]
Consider a common subsequence of strings $X$ and $Y$. Such subsequence uniquely determines two sets of indices $I$ and $J$ (as in definition \ref{def:cs}). If both $I$ and $J$ can be partitioned into families of sets of consecutive indices such that every set has a size of at least $k$, this subsequence is called a \textbf{k++ common subsequence}.
\end{definition}

\begin{definition}[LCSk++]
\text{$LCSk$++} of two strings $X$ and $Y$ is the length of their k++ common subsequence with maximal number of elements.
\end{definition}

\begin{definition}
We denote a substring of string $X$ starting at index $i$ and ending at index $j$ by $X_{i...j}$. If $i > j$ then $X_{i...j}$ denotes an empty string.
\end{definition}

\begin{example} 
 Consider the same strings as in Example \ref{example:lcsk.drawback}. Now $LCS3$++$(X,Y)$ equals 5 and $LCS3$++$(X,Z)$ equals 3, which reflects the fact that $X$ and $Y$ are more similar than $X$ and $Z$. That is because of the $k$-length substrings are allowed to overlap.
 \label{example:lcskpp}
\end{example}

\section{Computation}
\label{sec:computation}

\subsection{Basic dynamic programming}

The basic dynamic programming algorithm sequentially computes the values of $dp(i,j)=\text{$LCSk$++}(X_{0...i-1}, Y_{0...j-1})$ via the following recursive relation:

\begin{eqnarray}
  dp(i,j) & = & \max
  \left \{
    \begin{array}{ll}
      0 \\
      dp(i-1, j) & i \ge 1\\
      dp(i, j-1) & j \ge 1\\
      dp(i-q, j-q)+q & \text{for all $q \ge k$ s.t. $X_{i-q...i-1}=Y_{j-q...j-1}$}\\
    \end{array}
  \right .
\end{eqnarray}

The $2^{nd}$ and $3^{rd}$ terms in the above formula correspond to inheriting the $LCSk$++ value from previously computed values while the last term tries to extend the $\text{$LCSk$++}(X_{0...i-q-1}, Y_{0...j-q-1})$ with $X_{i-q...i-1}$ and $Y_{j-q...j-1}$ if they are equal. Those terms contribute $|X_{i-q...i-1}| = |Y_{j-q...j-1}| = q$ to the resulting length. A direct implementation of the above idea leads to an algorithm with time complexity $O(nm \cdot \min(n, m))$.

\subsection{Efficient algorithm}
\label{sec:efficient}

\begin{definition}[Match pair\footnote{Original definition given in \cite{lcsk}.}]
For a given strings $X$, $Y$ and integer $k \ge 1$ we define:
\begin{eqnarray}
kMatch(i,j) & = &
  \left \{
    \begin{array}{ll}
      1 & if X_{i+f}=Y_{j+f}, \text{for every }0 \le f \le k-1 \\
      0 & otherwise
    \end{array}
  \right .
\end{eqnarray}
If kMatch(i,j)=1, we call \emph{(i,j)} a \textbf{match pair}. In other words, kMatch(i,j)=1 when the substring of $A$ starting at $i$ and having an length of \textbf{exactly} $k$ is equal to the substring of $B$ starting at $j$ with the same length. $(i,j)$ is also called the \textbf{start} and $(i+k,j+k)$ is called the \textbf{end} of the match pair.
\end{definition}
When the number of match pairs $r$ is less than quadratic (consult Section \ref{sec:howtok} for the analysis), it is possible to compute $LCSk$++ efficiently. The time complexity of the algorithm we will describe in this section is $O(n + m + r \log r)$.\\
\\
For every match pair $P = (i_P, j_P)$ we use dynamic programming to compute $dp(P)=\text{$LCSk$++}(X_{0...i_P+k-1}, Y_{0...j_P+k-1})$, which represents the value of $LCSk$++ ending with $P$. The following definitions will be useful:

\begin{definition}[Precedence of match pairs]
  Let $P$=$(i_P, j_P)$ and $G$=$(i_G, j_G)$ be k-match pairs. Then $G$ \textbf{precedes} $P$ if $i_G+k \le i_P$ and $j_G+k \le j_P$. In other words, $G$ precedes $P$ if the end of G is on the upper left side of the start of P in the dynamic programming table (see Figure \ref{fig:kmatchpairs}).
\end{definition}

\begin{definition}[Continuation of match pairs]
  Let $P$=$(i_P, j_P)$ and $G$=$(i_G, j_G)$ be k-match pairs. Then $P$ \textbf{continues} $G$ if $i_P-j_P = i_G-j_G$ (i.e. they are on the same primary diagonal) and $i_P - i_G = 1$ ($P$ is only one down-right position from $G$, see Figure \ref{fig:kmatchpairs}).
\end{definition}

\begin{figure}[H]
\centering

\begin{tikzpicture}[scale=0.7]
\draw (0,-1) grid (11,6);
\node at (0.5, 6.5) {C};
\node at (1.5, 6.5) {T};
\node at (2.5, 6.5) {A};
\node at (3.5, 6.5) {T};
\node at (4.5, 6.5) {A};
\node at (5.5, 6.5) {G};
\node at (6.5, 6.5) {A};
\node at (7.5, 6.5) {G};
\node at (8.5, 6.5) {T};
\node at (9.5, 6.5) {A};
\node at (10.5, 6.5) {\$};

\node at (-0.5, 5.5) {A};
\node at (-0.5, 4.5) {T};
\node at (-0.5, 3.5) {T};
\node at (-0.5, 2.5) {A};
\node at (-0.5, 1.5) {T};
\node at (-0.5, 0.5) {G};
\node at (-0.5, -0.5) {\$};

\node at (1.5, 3.5) {a}; 
\draw (1.5,3.5) circle [radius=0.25]; 
\node at (3.5, 1.5) {a}; 
\draw (3.5,1.5) +(-0.25, -0.25) rectangle +(0.25, 0.25);

\node at (2.5, 2.5) {b};
\draw (2.5,2.5) circle [radius=0.25];
\node at (4.5, 0.5) {b};
\draw (4.5,0.5) +(-0.25, -0.25) rectangle +(0.25, 0.25);

\node at (2.5, 5.5) {c};
\draw (2.5,5.5) circle [radius=0.25];
\node at (4.5, 3.5) {c};
\draw (4.5,3.5) +(-0.25, -0.25) rectangle +(0.25, 0.25);

\node at (3.5, 3.5) {d};
\draw (3.5,3.5) circle [radius=0.25];
\node at (5.5, 1.5) {d};
\draw (5.5,1.5) +(-0.25, -0.25) rectangle +(0.25, 0.25);

\node at (8.5, 3.5) {e};
\draw (8.5,3.5) circle [radius=0.25];
\node at (10.5, 1.5) {e};
\draw (10.5,1.5) +(-0.25, -0.25) rectangle +(0.25, 0.25);
\end{tikzpicture}

\caption{$k = 2$; strings $X=ATTATG$ and $Y=CTATAGAGTA$ construct exactly five 2-match pairs denoted $a$ to $e$. Starts are represented by circles, ends are represented by squares. The following holds: ``$b$ continues $a$'', ``$c$ precedes $e$'', while the following does not hold: ``$a$ precedes $b$'', ``$c$ precedes $d$'', ``$a$ continues $b$''.}
\label{fig:kmatchpairs}
\end{figure}

We can express the $dp(P) = \text{$LCSk$++}(X_{0...i_P+k-1}, Y_{0...j_P+k-1})$ via the following formula which will be the basis for the efficient algorithm:
\begin{eqnarray}
  dp(P) & = & \max
  \left \{
    \begin{array}{ll}
      k \\
      \max_G dp(G) + k & \text{over all G preceding P}\\
      dp(G) + 1 & \text{if P continues G} 
    \end{array}
  \right .
\end{eqnarray}
In other words, a k-match pair $P$ can either start its own k-common subsequence, extend a k-common subsequence ending with a match $G$ such that $G$ precedes $P$ or extend a k-common subsequence ending with a match pair $G$ such that $P$ continues $G$. In the second case the k-common sequence is enlarged by $k$ (e.g. $c \to e$ in figure \ref{fig:kmatchpairs}), while in the latter it's enlarged by 1 (e.g. $a \to b$ in figure \ref{fig:kmatchpairs}).

\begin{algorithm}[h]
\begin{algorithmic}[1]  
\State $MaxColDp\gets$ 1D array filled with $n$ zeros
\State $MatchPairs \gets$ find all k-match pairs between $X$ and $Y$
\State $events \gets$ all starts and ends of $MatchPairs$ sorted in row-major order, if some start $S = (i_S, j_S)$ and some end $E=(i_E, j_E)$ share the same indices, $E$ should come first

\ForAll{$event \in events$}
  \If{$event$ is a start $P = (i_P, j_P)$}
    \State $dp(P)$ $\gets k+max_{x \in 0...j_P}$ $MaxColDp(x)$
  \ElsIf{$event$ is an end $P = (i_P+k, j_P+k)$}
	\If{$\exists G$ s.t. $P$ continues $G$}
	    \State $dp(P) \gets \max\{dp(P), dp(G)+1\}$
	\EndIf
    \State $MaxColDp(j_P + k) \gets \max \left \{ MaxColDp(j_P + k), dp(P) \right \} $
  \EndIf
\EndFor

\State \textbf{return} $\max_P dp(P)$
\end{algorithmic}
\caption{Efficient $LCSk$++ computation}
\label{alg:klcsfast}
\end{algorithm}

Algorithm \ref{alg:klcsfast} starts by extracting all of the $r$ match pairs on line 2. This can be done in two ways: we can employ a suffix array in the exact same manner as in \cite{lcskfast} to get the time complexity of $O(n + m + r)$. However, as $k$ is small in practice\footnote{Usually $k$ will be small enough such that every k-length substring can be perfectly hashed using 64 bits, see section \ref{sec:howtok} for details.} we can find all match pairs using a simple hash table in $O(n + m + kr) = O(n + m + r)$.

Line 3 creates events and sorts them, which ensures the correctness of the sweeping algorithm on lines 4-14. This can be accomplished in $O(r \log r)$ using a standard comparison based sorting algorithm. Line 8 can be implemented as a binary search over the $events$ array. If $MaxColDp$ is implemented as a Fenwick tree \cite{fenwick1994new}, the operations on lines 6 and 11 have a cost of $O(\log n)$ which implies that the sweep algorithm runs in $O(r \log n)$. Overall complexity is $O(m + n + r \log r)$\footnote{This is assuming that $r$ is at least as big as $n$. In the other case the correct complexity is $O(m + n + r \log r + r \log n$).}. The memory complexity is $O(n + m + r)$ because we only need the space to save the match pairs and the $MaxColDp$ structure. If we would like to reconstruct the sequence, the $dp$ array has to store $O(1)$ additional information per match pair: a pointer to the previous match pair in case some other match pair $G$ preceded $P$ or $P$ continued some $G$ in the optimal solution. We remark that by removing lines 8-10 this algorithm computes $LCSk$.

\section{How to choose $k$?}
\label{sec:howtok}

In this section we will analyze the performance of the $LCSk$++ measure on the following classification problem: given a pair of strings $X$ and $Y$, decide whether they are similar to each other. To formalize this problem, next sections proposes two simple random models: one for the unrelated pairs of strings and one for similar pairs. We use this setting to demonstrate the influence of the parameter $k$ to the performance and separability of $LCSk$++. 

On an efficiency note, it is natural to expect that the time complexity of the algorithm will decrease as the size of the alphabet increases (due to the diminishing number of match pairs). The interesting questions thus emerge when the size of the alphabet is small and the lengths of the strings are large. Such a setting is naturally found in DNA sequence alignment, which is an inspiration for the presented model.

\subsection{Similarity model}

We model a pair of {\bfseries unrelated strings} $X^{(n)}$ and $Y^{(n)}$ of length $n$ as random strings over the alphabet $\{A, C, G, T \}$ where each character has a known and mutually independent probability of appearing. This construction yields a constant $e_{unrelated}$ defined as $e_{unrelated} = P[ X_{i}^{(n)} \neq Y_{i}^{(n)}]$.

The model for {\bfseries similar pairs} depends on a fixed parameter $e_{similar} < e_{unrelated}$. Pairs of strings are generated in a way that one of them is created randomly over $\{ A, C, G, T \}$ using the same distributions as above, and the other one is its mutated copy satisfying restriction $P[X_{i}^{(n)} \neq Y_{i}^{(n)}] = e_{similar}$. The model reflects simplified evolutionary mutations (only substitutions are considered).

\begin{figure}[ht]
	\centering
	\begin{subfigure}[b]{0.49\textwidth}
                \centering
                \includegraphics[trim=1.2cm 0cm 2.5cm 0cm, clip=true, width=\textwidth]{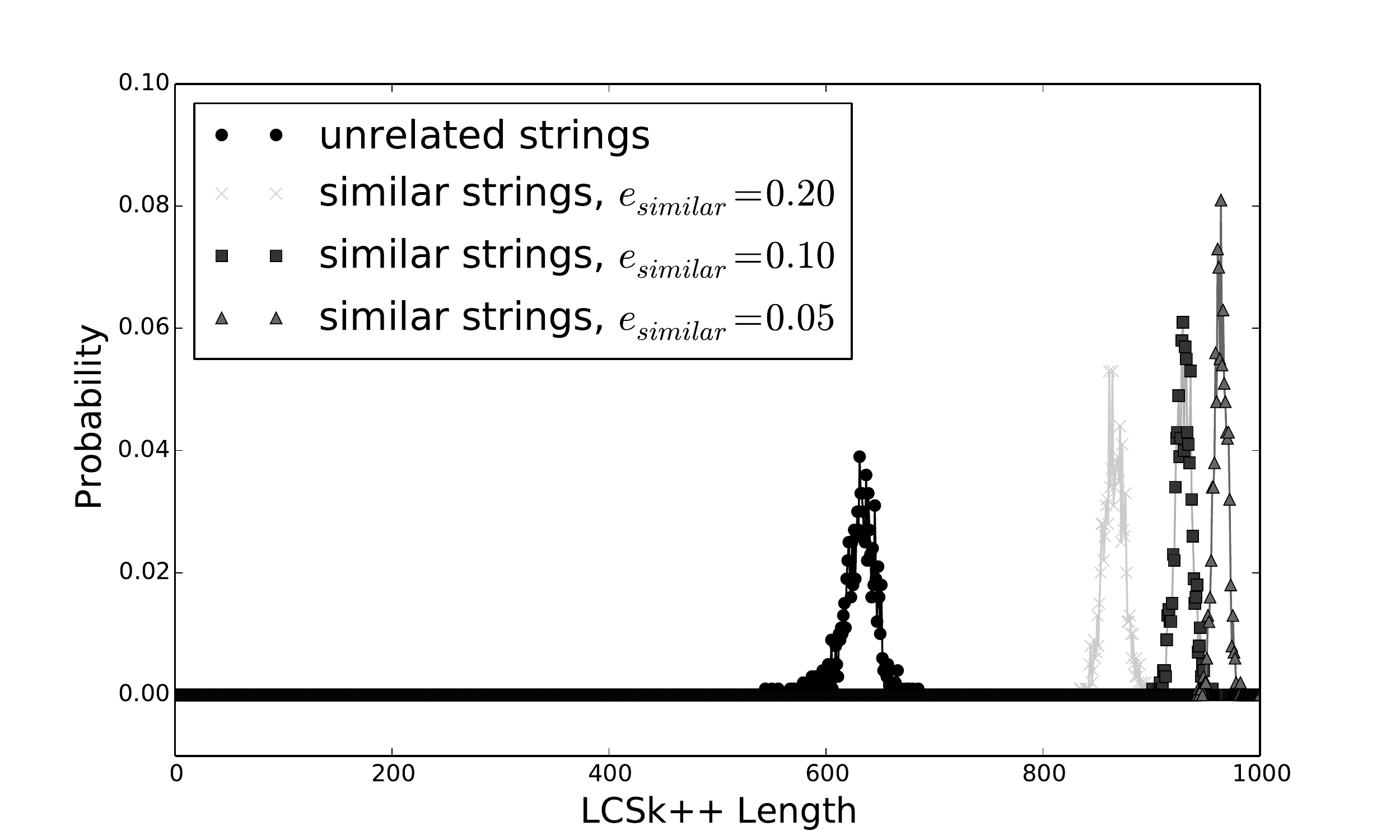}
                \caption{$k = 1$}
                \label{fig:klcs_n1000_seedlen1}
	\end{subfigure}
	\begin{subfigure}[b]{0.49\textwidth}
                \centering
                \includegraphics[trim=1.2cm 0cm 2.5cm 0cm, clip=true, width=\textwidth]{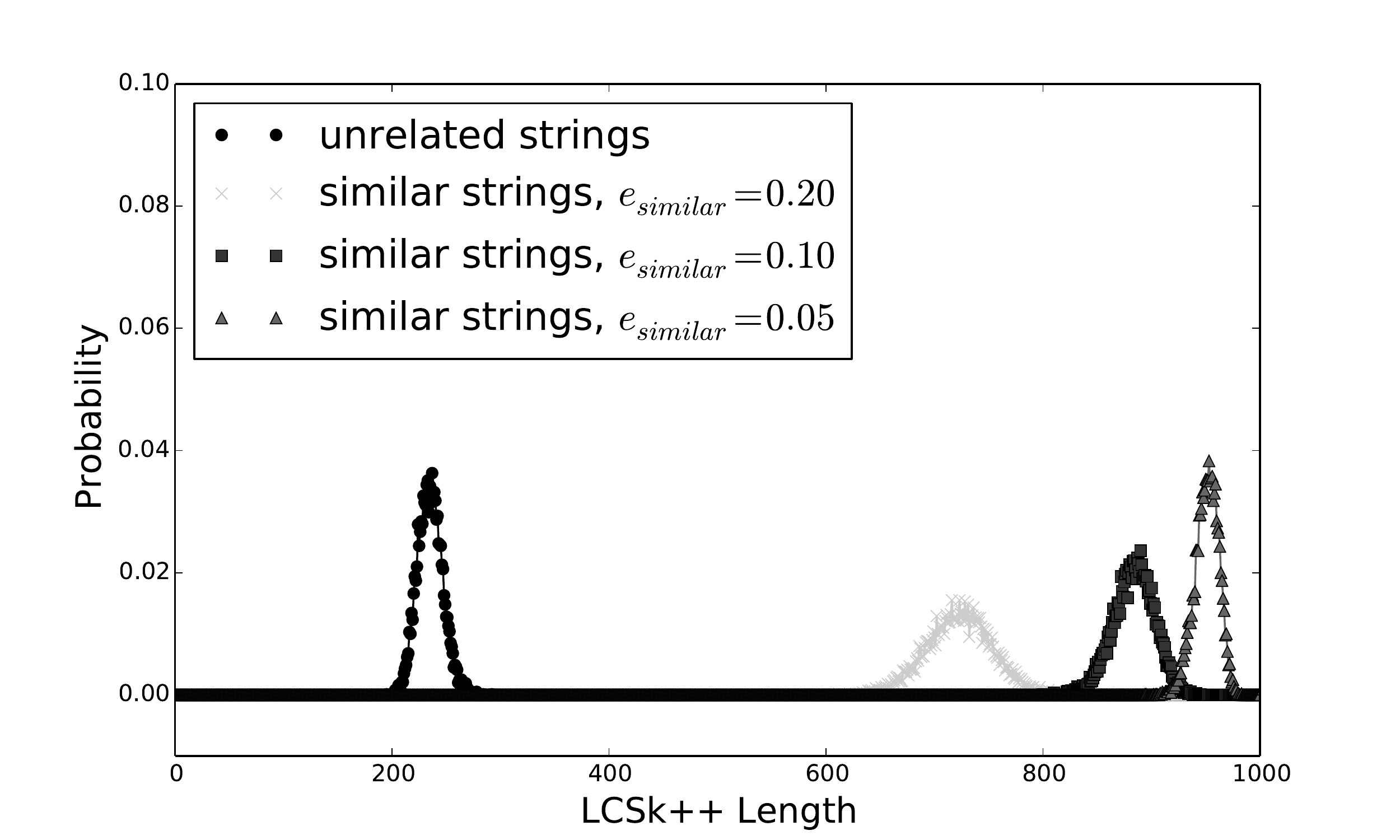}
                \caption{$k = 5$}
                \label{fig:klcs_n1000_seedlen5}
	\end{subfigure}

	\centering
	\begin{subfigure}[b]{0.49\textwidth}
                \centering
                \includegraphics[trim=1.2cm 0cm 2.5cm 0cm, clip=true, width=\textwidth]{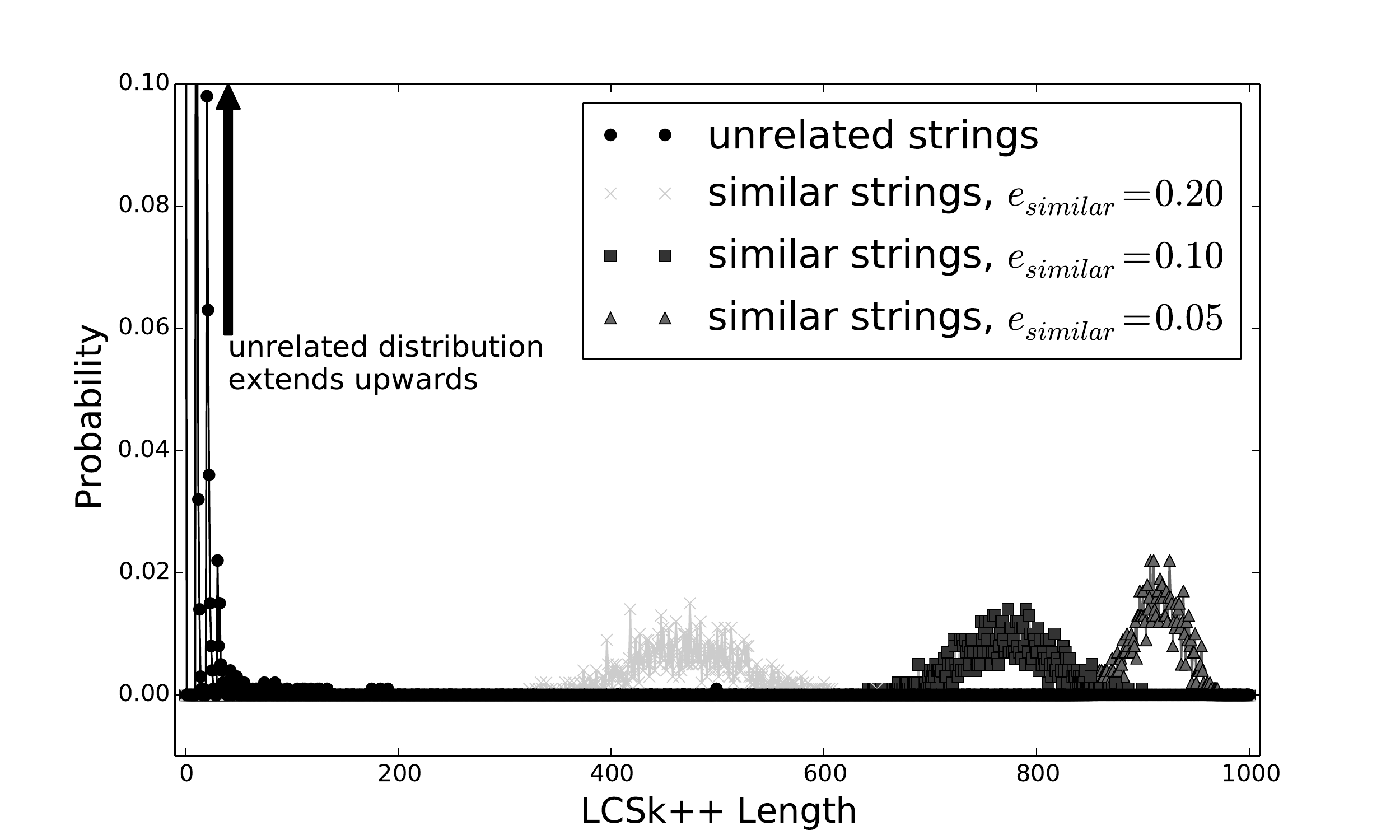}
                \caption{$k = 10$}
                \label{fig:klcs_n1000_seedlen10}
	\end{subfigure}
	\begin{subfigure}[b]{0.49\textwidth}
                \centering
                \includegraphics[trim=1.2cm 0cm 2.5cm 0cm, clip=true, width=\textwidth]{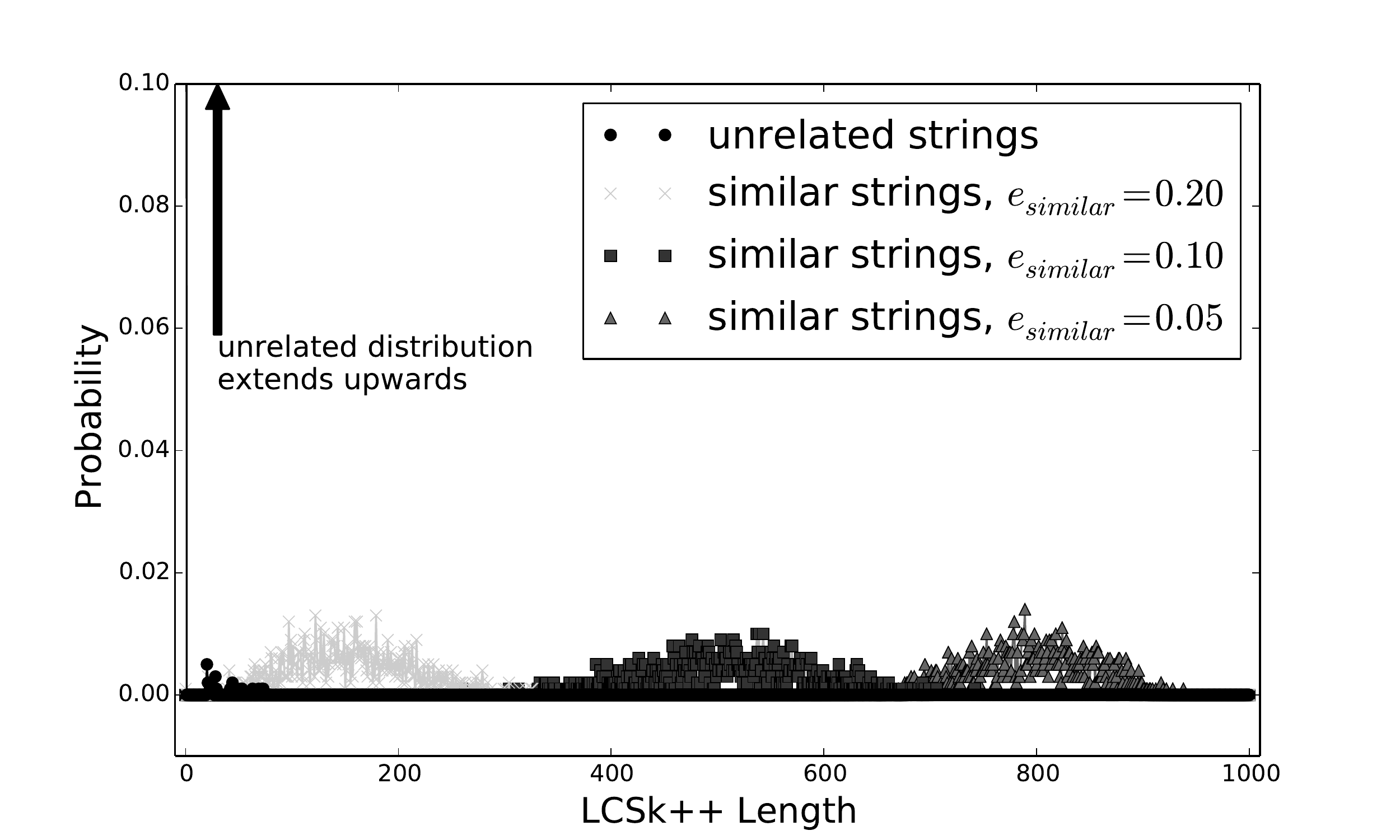}
                \caption{$k = 20$}
                \label{fig:klcs_n1000_seedlen20}
	\end{subfigure}
  \caption{\textbf{$LCSk$++ distributions} for strings of length 1000 and several values of $k$. It is clear that increasing $k$ lowers the average score that string pairs achieve and increases the deviation. For our similarity model, the boundary of good separability is reached by setting $k$ to around 20. DNA substrings of length 20 can be perfectly hashed in a 64-bit integer. That is useful for implementation because the step of finding all the match pairs can be implemented with a simple hash table.}
  \label{fig:klcs.distributions}
\end{figure}

In order to distinguish between two classes, we would expect the following to hold (see Figure \ref{fig:klcs.distributions} for confirmation of this claim):
\begin{equation}
\label{eqn:E-classifier}
\frac{E[\text{LCSk++}(X^{(n)}_{similar}, Y^{(n)}_{similar})]}{n} > \frac{E[\text{LCSk++}(X^{(n)}_{unrelated}, Y^{(n)}_{unrelated})]}{n}
\end{equation}
Computing the asymptotic behavior in Eq. \ref{eqn:E-classifier} is still an open problem \cite{bundschuh.sim,Kiwi:2009:SRC:1552058.1552061}, though the limit does exist - it can be shown by following the argument for the expectation of LCS for random strings by Chv\'{a}tal and Sankoff in \cite{chvatalsankoff}. We compute them by a Monte Carlo simulation.\\
Unpublished work of Rabinovitch\cite{rabinovitch} experimentally demonstrates that the value of $\frac{E[LCS(X^{(n)}, Y^{(n)})]}{n}$ increases with $n$, eventually hitting the limit. The value of $\frac{StdDev[LCS(X^{(n)}, Y^{(n)})]}{n}$ decreases with $n$. Experiments presented in Table \ref{tbl:klcs_simulation_acc} confirm this behavior for $LCSk$++. Effectively this means that the separability gets better as $n$ increases.

\subsection{Expected value of match pairs $r$}
Let $X$, $Y$ be either unrelated or similar pair of strings of length $m, n$ constructed over the alphabet \{A, C, G, T\} with a priori distributions $p_A, p_C, p_G, p_T$. Then $S := E[X_i = Y_j] = (p_A^2 + p_C^2 + p_G^2 + p_T^2)$ for all $i \neq j$ (this restriction is needed to cover both models with this proof). Expected number of match pairs then equals:

\begin{eqnarray}
E[r] & = & E[\text{\# of pairs (i,j) such that } X_{i..i+k-1} = Y_{j..j+k-1}] \nonumber \\
& = & E\left[\sum_{i=0}^{n-k} \sum_{j=0}^{m-k} \mathbbm{1}[X_{i..i+k-1} = Y_{j..j+k-1}]\right] \, \nonumber \\
& = & E\left[\sum_{i=0}^{n-k} \sum_{j=0}^{m-k} \mathbbm{1}[i = j] \mathbbm{1}[X_{i..i+k-1} = Y_{j..j+k-1}]\right] \, + \nonumber \\
& & E\left[\sum_{i=0}^{n-k} \sum_{j=0}^{m-k} \mathbbm{1}[i \neq j] \mathbbm{1}[X_{i..i+k-1} = Y_{j..j+k-1}]\right] \nonumber \\
& = & O(n + m) + \sum_{i=0}^{n-k} \sum_{j=0}^{m-k} E\left[\mathbbm{1}[i \neq j] \mathbbm{1}[X_{i..i+k-1} = Y_{j..j+k-1}]\right] \nonumber \\
& = & O(n + m + nmS^k)
\end{eqnarray}

\begin{corollary}
By choosing $k_{fast} = \log_{1/S} \frac{nm}{n+m}$ it follows that $E[r] = O(n + m)$, so the expected complexity of the whole algorithm is $O((n + m)\log(n + m))$.
\end{corollary}

\begin{corollary}
For uniformly distributed alphabets, the expected number of match pairs drops as the size of the alphabet increases. That implies that the bigger the alphabet is, the smaller $k$ is needed for the $LCSk$++ computation to run efficiently.
\end{corollary}

\begin{table}[ht]
  \centering
  \caption{\textbf{Expected values and standard deviations of the $LCSk$++ distribution} for strings of length 100, 1000 and 10000, $k$ of 10 and 20, and for unrelated and similar pairs (with various $e_{similar}$).\\}

	\begin{subfigure}[b]{0.49\textwidth}
    \centering
    \resizebox{\linewidth}{!}{
    \begin{tabu}{l @{\hspace{1em}} l @{\hspace{2em}} l @{\hspace{1em}} c @{\hspace{1em}} c} \toprule
      \bfseries k & \bfseries string length & \bfseries error & $\bm{\frac{E[LCSk++]}{n}}$ & \bfseries $\bm{\frac{Std. Dev.}{n}}$ \\
      \midrule

      10 & 1000 & unrelated & 0.015 & 0.024 \\
      & & 0.20 & 0.470 & 0.051 \\
      & & 0.10 & 0.770 & 0.041 \\
      & & 0.05 & 0.911 & 0.025 \\[.5ex]
      & 10000 & unrelated & 0.032 & 0.015 \\
      & & 0.20 & 0.471 & 0.017 \\
      & & 0.10 & 0.772 & 0.014 \\
      & & 0.05 & 0.914 & 0.008 \\[.5ex]
      & 100000 & unrelated & 0.041 & 0.015 \\
      & & 0.20 & 0.471 & 0.006 \\
      & & 0.10 & 0.772 & 0.005 \\
      & & 0.05 & 0.914 & 0.003 \\

      \bottomrule
    \end{tabu}    
  }
	\end{subfigure}
	\begin{subfigure}[b]{0.49\textwidth}    
    \centering
    \resizebox{\linewidth}{!}{
    \begin{tabu}{l @{\hspace{1em}} l @{\hspace{2em}} l @{\hspace{1em}} c @{\hspace{1em}} c} \toprule
      \bfseries k & \bfseries string length & \bfseries error & $\bm{\frac{E[LCSk++]}{n}}$ & \bfseries $\bm{\frac{Std. Dev.}{n}}$ \\
      \midrule

      20 & 1000 & unrelated & 0.001 & 0.006 \\
      & & 0.20 & 0.154 & 0.057 \\
      & & 0.10 & 0.512 & 0.075 \\
      & & 0.05 & 0.793 & 0.058 \\[.5ex]
      & 10000 & unrelated & 0.006 & 0.009 \\
      & & 0.20 & 0.154 & 0.018 \\
      & & 0.10 & 0.516 & 0.025 \\
      & & 0.05 & 0.801 & 0.018 \\[.5ex]
      & 100000 & unrelated & 0.011 & 0.008 \\
      & & 0.20 & 0.154 & 0.006 \\
      & & 0.10 & 0.516 & 0.008 \\
      & & 0.05 & 0.801 & 0.006 \\
    
      \bottomrule
    \end{tabu}    
  }
	\end{subfigure}


\label{tbl:klcs_simulation_acc}
\end{table}

\section{Conclusions}

In this paper we defined $LCSk$++, a similarity metric for long strings and we proposed an efficient algorithm for its computation. $LCSk$++ is a natural extension over the previous $LCSk$ metric \cite{lcsk}, with improved sensitivity. The only parameter $k$ gives a flexible tradeoff between computational efficiency and sensitivity. Assuming strings $X$ and $Y$ follow a realistic random model, we presented a $O((|X| + |Y|)\log(|X| + |Y|))$ algorithm for computing the metric. Because of dependence on only a simple Fenwick tree data structure, the implementation is almost straightforward, making $LCSk$++ attractive for usage in practice.

\subsubsection{Acknowledgments.}
The authors would like to thank Mario Lu\v{c}i\v{c} for valuable comments on the manuscript. This work was supported by the University of Zagreb under grant "Efficient algorithms for data processing in bioinformatics" and the Croatian academy of sciences and arts.

\bibliography{refs}
\bibliographystyle{ieeetr}

\end{document}